\documentclass{ws-ijmpa}
\usepackage[super,compress]{cite}
\usepackage{graphicx}

\begin{document}
\markboth{R.~Reifarth}{Treatment of isomers in nucleosynthesis codes}

%
\catchline{}{}{}{}{}
%

\title{Treatment of isomers in nucleosynthesis codes}

\author{Ren\'e~Reifarth, Stefan Fiebiger, Kathrin G\"obel,  Tanja Heftrich, 
Tanja~Kausch, Christoph~K\"oppchen, Deniz~Kurtulgil, Christoph Langer, Benedikt Thomas, Mario Weigand}

\address{Goethe-Universit\"{a}t Frankfurt am Main\\
 Max-von-Laue-Str.1 \\
 Frankfurt am Main, 60438, Germany\\
 reifarth@physik.uni-frankfurt.de}

\maketitle{}

\today\quad\currenttime

\begin{history}
\received{Day Month Year}
\revised{Day Month Year}
\end{history}

\begin{abstract}
The decay properties of long-lived excited states (isomers) can have a significant impact on the destruction channels of isotopes under stellar conditions. In sufficiently hot environments, the population of isomers can be altered via thermal excitation or de-excitation. If the corresponding lifetimes are of the same order of magnitude as the typical time scales of the environment, the isomers have to be the treated explicitly. We present a general approach to the treatment of isomers in stellar nucleosynthesis codes and discuss a few illustrative examples. The corresponding code is available online at http://exp-astro.de/isomers/

\end{abstract}

\ccode{PACS numbers:}


\section{Introduction}
Isomers are excited states with significantly longer half lives than the typical excited states of nuclei. This implies a different treatment in all applications where the conditions change explicitly as a function of time. The exact separation depends on the context. When a coincidence requirement in a nuclear physics experiment is discussed, even half lives of nanoseconds have to be treated as isomers. The typical time scales in stellar nucleosynthesis codes, however, range from milliseconds \cite{FML06,ArT13} during explosions to millions of years during stable burning phases in stars \cite{RLK14}. The correct treatment of the abundances of isomers is of particular importance if the destruction rates are different for the different isomers. Within this article, we will only consider $\beta$-decays as destruction rates. The generalization to more destruction channels like neutron capture, charged particle reactions etc. is obvious.  

\section{Terrestrial conditions}

\begin{figure}[h]
      \begin{center}
       \includegraphics[width=.7\textwidth]{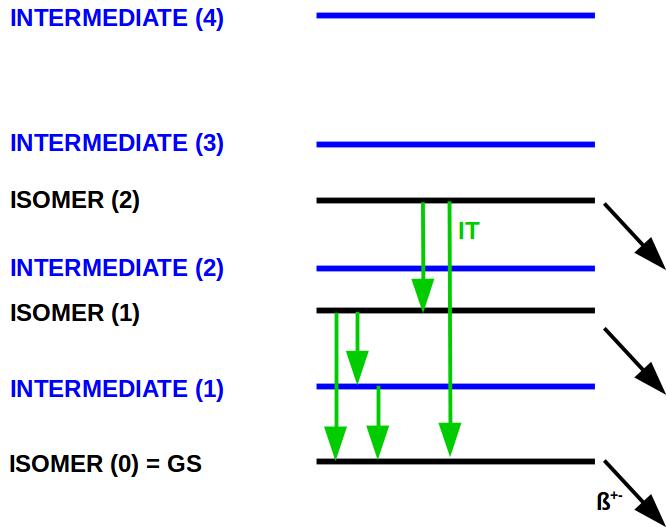}
       \caption{Situation under terrestrial conditions. The levels can only decay via internal
                transition or $\beta$-decays. Higher lying levels will therefore not be populated.
               }\label{all_terrestrial}
      \end{center}
      \end{figure}

Isomers, ground state and intermediate levels under terrestrial (cold) conditions are schematically shown in Fig.~\ref {all_terrestrial}. Isomers will either $\beta$-decay or de-excite via internal transition. The internal transition can lead to short-lived excited states or to other isomers including the ground state. Intermediate states are only involved if they are populated through internal transitions. 

The terrestrial properties of each state (including intermediate states) can be summarized as:

\begin{itemize}
	\item{$E_i$ is the excitation energy $(i<j \iff E_i < E_j)$}
	\item{$g_i = 2 l_i +1$ is the degeneracy of level $i$ with the angular momentum $l_i$}
	\item{$\lambda_i = 1/\tau_i = \ln{2} / t_i^{1/2}$ is the terrestrial decay constant}
	\item{$\beta_i$ is the $\beta$-branching, this can also be $\alpha$-decay, neutron capture or similar reactions changing $A$ or $Z$}
	\item{$\eta_{i \rightarrow j}$ is the internal decay branchings (IT), from state $i$ to state $j$,
	  ($i > j$). Important: $\eta_{i \rightarrow j} = 0,~\forall~ i \le j$ }
\end{itemize}

Conventions for each state $i$:

Normalization:
\begin{equation}
  \beta_i + \sum_{j<i}{  \eta_{i \rightarrow j} } =1
\end{equation}

$\beta$-decay constants:
\begin{equation}
  \lambda_{\beta_i}:=\beta_i \lambda_i = \beta_i/\tau_i
\end{equation}

Transition constants between individual states:
\begin{equation}
  \lambda_{i \rightarrow j}:=\eta_{i \rightarrow j} \lambda_i = \eta_{i \rightarrow j}/\tau_i~~;~~\lambda_{i \rightarrow j} = 0,~\forall~ i \le j
\end{equation}

Following from the equations above:
\begin{equation}
  \lambda_i = \lambda_{\beta_i} + \sum_{j<i} {\lambda_{i \rightarrow j}} 
\end{equation}

\section{Stellar conditions}

\begin{figure}[h]
      \begin{center}
       \includegraphics[width=.7\textwidth]{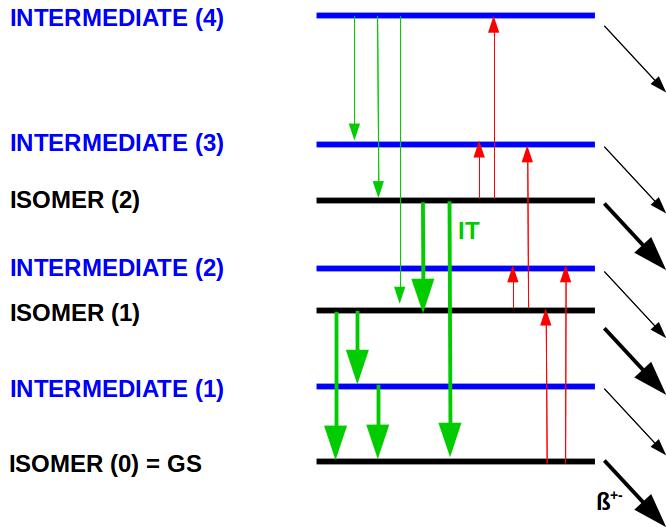}
       \caption{Situation under stellar conditions. In general, all states can now be thermally excited 
               (red). Higher-lying levels can be populated and can in turn decay to other
               low-lying levels. This mechanism to thermally couples long-lived states.
               }\label{all_stellar}
      \end{center}
      \end{figure}

Under stellar conditions, in contrast to terrestrial conditions, higher-lying 
levels can be populated from low-lying levels via thermal excitation. 
These excitations are caused by photons. The number of these photons depends strongly 
on the temperature $T$ of the stellar environment, see Fig.~\ref{all_stellar}. In principle, these high-lying states can even be above the particle separation threshold, which leads to additional destruction channels  \cite{PGR16}.

The corresponding excitation rate $\lambda^T_{j \rightarrow i}$, ($j < i$), can only the states $j$ and $i$ exist but no $\beta$-branch. 
A thermal equilibrium is reached for the number of atoms in the corresponding states $N_j$ and $N_i$:

\begin{equation}
  \frac{N_i}{N_j}=\frac{g_i}{g_j} \mbox{e}^{-\frac{E_i-E_j}{kT}}
\end{equation}

(Maxwell-Boltzmann factor) and

\begin{equation}
  \lambda_{i \rightarrow j} N_i = \lambda^T_{j \rightarrow i} N_j
\end{equation}

(equilibrium condition), hence:

\begin{equation}
  \lambda^T_{j \rightarrow i} = \lambda_{i \rightarrow j} \frac{g_i}{g_j} \mbox{e}^{\frac{E_j-E_i}{kT}}
\end{equation}

\begin{equation}
  \lambda^T_{i} := \sum_{j>i}{\lambda^T_{i \rightarrow j} }
\end{equation}

Similar to the definitions above, the stellar properties of each state 
(including intermediate states) are:

\begin{itemize}
	\item{$\lambda_{\beta_i}^\star = \lambda_{\beta_i}$ is the stellar $\beta$-decay rate, we will ignore bound-state $\beta$-decays 
	      for now \cite{TaY87}. This feature will be included in an updated version. Bound-state $\beta$-decays are is only important for states with very low excitation energy, typically only the ground state.}

	\item{$\lambda_i^\star$ is the stellar decay constant}

      \begin{eqnarray}
        \lambda_i^\star & = & \lambda_i + \lambda^T_{i} \\
                        & = & \lambda_i + \sum_{j>i}{\lambda^T_{i \rightarrow j} } \\
                        & = & \lambda_i + \sum_{j>i}{\lambda_{j \rightarrow i} \frac{g_j}{g_i} \mbox{e}^{-\frac{E_j-E_i}{kT}} } \\
                        & = & \lambda_i + \sum_{j>i}{\eta_{j \rightarrow i} \lambda_j \frac{g_j}{g_i} \mbox{e}^{-\frac{E_j-E_i}{kT}} } \\
                        & = & \lambda_{\beta_i} + \sum_{j<i} {\lambda_{i \rightarrow j}}  + \sum_{j>i}{\eta_{j \rightarrow i} \lambda_j \frac{g_j}{g_i} \mbox{e}^{-\frac{E_j-E_i}{kT}} } 
        \end{eqnarray}

\end{itemize}

\section{Effective stellar rates}

The treatment of isomers in nucleosynthesis codes is very often simplified: at temperatures below a critical value the species are treated as under terrestrial  conditions and above that, the species are treated as completely (instantly) coupled with the same average $\beta$-decay constant for all isomers. Apart from the discontinuity, this approach neglects the fact that it takes time to reach the thermal equilibrium between the isomeric states. A clean approach would be to treat all isomeric states explicitly, which means follow and record their abundance distributions separately. 

The goal is now to neglect all states in the explicit treatment where $\lambda^T_{i} << \lambda_{i}$. 
These are states with very short terrestrial half live. Such states fulfill the following important 
conditions:

\begin{itemize}
  \item{$\lambda^T_{i}$ can and will be neglected.}
  \item{None of the rates destroying these states depends on the stellar conditions. 
        Therefore all the branchings are the same as under terrestrial conditions, the same as
        defined in the corresponding input files.}
  \item{The decays are always much faster than the changes of the astrophysical conditions. 
        It is therefore safe to assume that the abundance of these states reaches the
        equilibrium of production and destruction very fast. Their destruction rate is always 
        the same as the production rate, which determines their abundance.}      
\end{itemize}

\begin{figure}[h]
      \begin{center}
       \includegraphics[width=.7\textwidth]{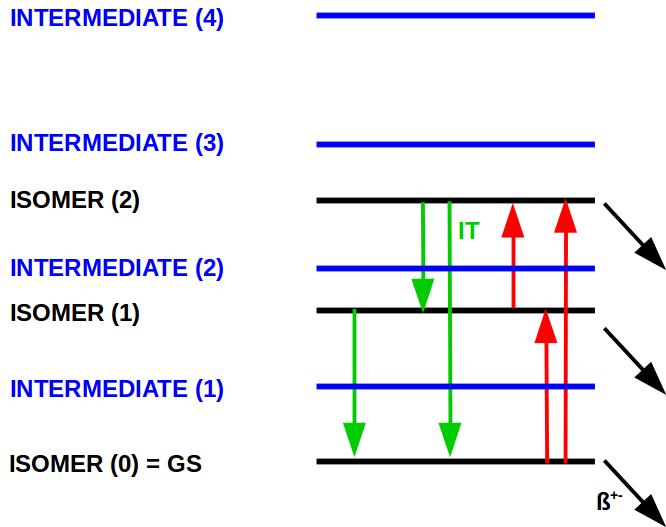}
       \caption{Simplified situation under stellar conditions. In addition to the transitions
                possible under terrestrial conditions (Fig.~\ref{all_terrestrial}), the 
                long-lived states are now
                thermally coupled. These transitions (red) depend strongly 
                on the temperature of the stellar environment.
               }\label{simplified_stellar}
      \end{center}
      \end{figure}

In other words: only the abundance of long-lived or stable states 
should be treated explicitly in the nucleosynthesis code - these are isomers and ground states. 
Therefore, effective rates are necessary, which 
couple the different long-lived states with each other and consider the possible $\beta$-decays via 
short-lived intermediate states correctly, see Fig.~\ref{simplified_stellar}. In order to make this 
distinction clear, all effective decay constants will have the symbol $\Lambda$ instead of $\lambda$.

We will now derive 

\begin{itemize}
  \item{$\Lambda_{i \rightarrow j}(T)$ is the effective coupling constant from the long-lived state $i$ to the 
           long-lived state $j$ considering all intermediate states. Under terrestrial conditions
           ($T\approx 0$), one finds:}
    \begin{itemize}
      \item{ $\Lambda_{i \rightarrow j}(0) = 0,~\forall ~ i \le j$ and }
      \item{ $\Lambda_{i \rightarrow j}(0) = \lambda_{i \rightarrow j} + \lambda^{\mathrm{inter}(k)}_{i \rightarrow j},~\forall ~i > j$. }
    \end{itemize}  
  \item $\lambda^{\mathrm{inter}(k)}_{m \rightarrow n}$ is the decay constant from state $m$ to state $n$ (long- or short-lived) via $k$ intermediate levels excluding the direct transition. No long-lived level is considered between $m$ and $n$ - the transition would stop there. If there is no intermediate level between $m$ und $n$, $\lambda^\mathrm{inter(0)}_{m \rightarrow n} = 0$. If we consider only the lowest intermediate level $(1^\star)$ between $m$ und $n$, we find:
    \begin{equation}\label{eq:inter1}
      \lambda^\mathrm{inter(1)}_{m \rightarrow n} = \lambda_{m \rightarrow 1^\star} \eta_{1^\star \rightarrow n}
    \end{equation}
   The lowest two intermediate levels $(1^\star)$ and $(2^\star)$ with $E_1^\star < E_2^\star$ between $m$ und $n$ would result in:
    \begin{eqnarray*}
      \lambda^\mathrm{inter(2)}_{m \rightarrow n} & = & \lambda_{m \rightarrow 2^\star} \frac{\lambda_{2^\star \rightarrow n}+\lambda^\mathrm{inter(1)}_{2^\star \rightarrow n}}{\lambda_{2^\star}} + \lambda_{m \rightarrow 1^\star} \eta_{1^\star \rightarrow n} \\
                                                  & = & \lambda_{m \rightarrow 2^\star} \frac{\lambda_{2^\star \rightarrow n}+\lambda^\mathrm{inter(1)}_{2^\star \rightarrow n}}{\lambda_{2^\star}} + \lambda^\mathrm{inter(1)}_{m \rightarrow n}
    \end{eqnarray*}
    Adding the $k$-th intermediate level with $E_{k-1} < E_k$ between $m$ und $n$ results in:
    \begin{equation}\label{eq:inter2}
      \lambda^{\mathrm{inter}(k)}_{m \rightarrow n} = \lambda_{m \rightarrow k} \frac{\lambda_{k \rightarrow n}+\lambda^{\mathrm{inter}(k-1)}_{k \rightarrow n}}{\lambda_{k}} + \lambda^{\mathrm{inter}(k-1)}_{m \rightarrow n}
    \end{equation}
    Since $\lambda_{k \rightarrow n} = 0$ for $E_k<E_n$ and $\lambda_{m \rightarrow k} = 0$ for $E_k > E_m$, we can drop the condition that the intermediate levels have to be between $m$ and $n$. We can simply use equations (\ref{eq:inter1}) (\ref{eq:inter2}) running over all short-lived levels.
 \item $\Lambda_{\beta_i}(T)$ is the effective $\beta$-decay constant of the long-lived state $i$ 
                considering all intermediate states. 
                
\end{itemize}

In order to quantify the population
of a long-lived state $j$ from another long-lived state $i$, one simply has to add up all the branches
leading from state $i$ to state $j$ via intermediate states. If the branch ends up in a $\beta$-decay 
or a long-lived state besides $j$, the branch does not contribute to $\Lambda_{i \rightarrow j}(T)$.

Let $i,j \in [1,n]$ be long-lived states and $k,l \in [1^\star,m]$ 
be intermediate states (number-indexes referring to intermediate states are marked with a $^\star$). 
May $i<j$ for now, then:

\begin{eqnarray*}
  \Lambda_{i \rightarrow j}(T) = &   &  \lambda^T_{i \rightarrow j}  									\\
                             & + &  \lambda^T_{i \rightarrow m} \frac{\lambda_{m \rightarrow j}+\lambda^{\mathrm{inter}(m-1)}_{m \rightarrow j}}{\lambda_{m}}        \\
                             & + &  \lambda^T_{i \rightarrow m-1} \frac{\lambda_{m-1 \rightarrow j}+\lambda^{\mathrm{inter}(m-2)}_{m-1 \rightarrow j}}{\lambda_{m-1}}        \\
                             & + &  ...  \\
                             & + &  \lambda^T_{i \rightarrow {2^\star}} \frac{\lambda_{2^\star \rightarrow j}+\lambda^{\mathrm{inter}(1)}_{2^\star \rightarrow j}}{\lambda_{2^\star}}	\\
                             & + &  \lambda^T_{i \rightarrow 1^\star} \eta_{1^\star \rightarrow j}  	
\end{eqnarray*}

hence,

\begin{eqnarray*}
  \Lambda_{i \rightarrow j}(T) &  = &  \lambda^T_{i \rightarrow j} + 
                                       \sum_{l=2}^{m}{\lambda^T_{i \rightarrow l} \frac{\lambda_{l \rightarrow j}+\lambda^{\mathrm{inter}(l-1)}_{l \rightarrow j}}{\lambda_{l}} } + \lambda^T_{i \rightarrow 1^\star} \eta_{1^\star \rightarrow j} 									\\
                       &  = &  \lambda_{j \rightarrow i} \frac{g_j}{g_i} \mbox{e}^{\frac{E_i-E_j}{kT}} + 
                               \sum_{l=2}^{m}{\lambda_{l \rightarrow i} \frac{g_l}{g_i} \mbox{e}^{\frac{E_i-E_l}{kT}} \frac{\lambda_{l \rightarrow j}+\lambda^{\mathrm{inter}(l-1)}_{l \rightarrow j}}{\lambda_{l}} } + 
                               \lambda_{{1^\star} \rightarrow i} \frac{g_{1^\star}}{g_i} \mbox{e}^{\frac{E_i-E_{1^\star}}{kT}} \eta_{1^\star \rightarrow j}		
\end{eqnarray*}

Only the very first term changes, if $j<i$:

\begin{eqnarray*}
  \Lambda_{i \rightarrow j}(T) &  = &  \lambda_{i \rightarrow j} + \lambda^{\mathrm{inter}(m)}_{i \rightarrow j} + 
                                       \sum_{l=2}^{m}{\lambda^T_{i \rightarrow l} \frac{\lambda_{l \rightarrow j}+\lambda^{\mathrm{inter}(l-1)}_{l \rightarrow j}}{\lambda_{l}} } + \lambda^T_{i \rightarrow 1^\star} \eta_{1^\star \rightarrow j} 									\\
                       &  = &  \lambda_{i \rightarrow j} + \lambda^{\mathrm{inter}(k)}_{i \rightarrow j} + 
                               \sum_{l=2}^{m}{\lambda_{l \rightarrow i} \frac{g_l}{g_i} \mbox{e}^{\frac{E_i-E_l}{kT}} \frac{\lambda_{l \rightarrow j}+\lambda^{\mathrm{inter}(l-1)}_{l \rightarrow j}}{\lambda_{l}} } + 
                               \lambda_{{1^\star} \rightarrow i} \frac{g_{1^\star}}{g_i} \mbox{e}^{\frac{E_i-E_{1^\star}}{kT}} \eta_{1^\star \rightarrow j}		
\end{eqnarray*}

The effective $\beta$-decay rate of state $i$ can be derived via:

\begin{eqnarray*}
  \Lambda_{\beta_i}(T) = &   &  \lambda_{\beta_{i}}  									\\
                         & + &  \sum_{l=1}^{m}{ \left( \lambda_{i \rightarrow l} + \lambda^{\mathrm{inter}(m)}_{i \rightarrow l} \right) \beta_{l}}	\\
                         & + &  \sum_{k=1}^{m} \lambda^T_{i \rightarrow k} \left( \beta_{k} + \sum_{l=1}^{k}{\frac{\lambda_{k \rightarrow l} + \lambda^{\mathrm{inter}(k-1)}_{k \rightarrow l}} {\lambda_k} \beta_l } \right)  	
\end{eqnarray*}

hence,

\begin{eqnarray*}
  \Lambda_{\beta_i}(T) = &   &  \lambda_{\beta_{i}}  									\\
                         & + &  \sum_{l=1}^{m}{ \left( \lambda_{i \rightarrow l} + \lambda^{\mathrm{inter}(m)}_{i \rightarrow l} \right) \beta_{l}}	\\
                         & + &  \sum_{k=1}^{m} \lambda_{k \rightarrow i} \frac{g_k}{g_i} \mbox{e}^{\frac{E_i-E_k}{kT}} \left( \beta_{k} + \sum_{l=1}^{k}{\frac{\lambda_{k \rightarrow l} + \lambda^{\mathrm{inter}(k-1)}_{k \rightarrow l}} {\lambda_k} \beta_l } \right)  	
\end{eqnarray*}

Now all effective rates are expressed in terms of the properties of the states under terrestrial conditions. 

\section{Online tools}

The terrestrial properties of the unstable isotopes and their excited states can easily be collected in simply-structured input-files. Usually the number of long-lived and short-lived states are small, hence, the general equations will become much simpler. But a generalized code was developed, which calculates the rates for any given number of states based on these input-files. 

\begin{figure}[h]
      \begin{center}
       \includegraphics[width=.9\textwidth]{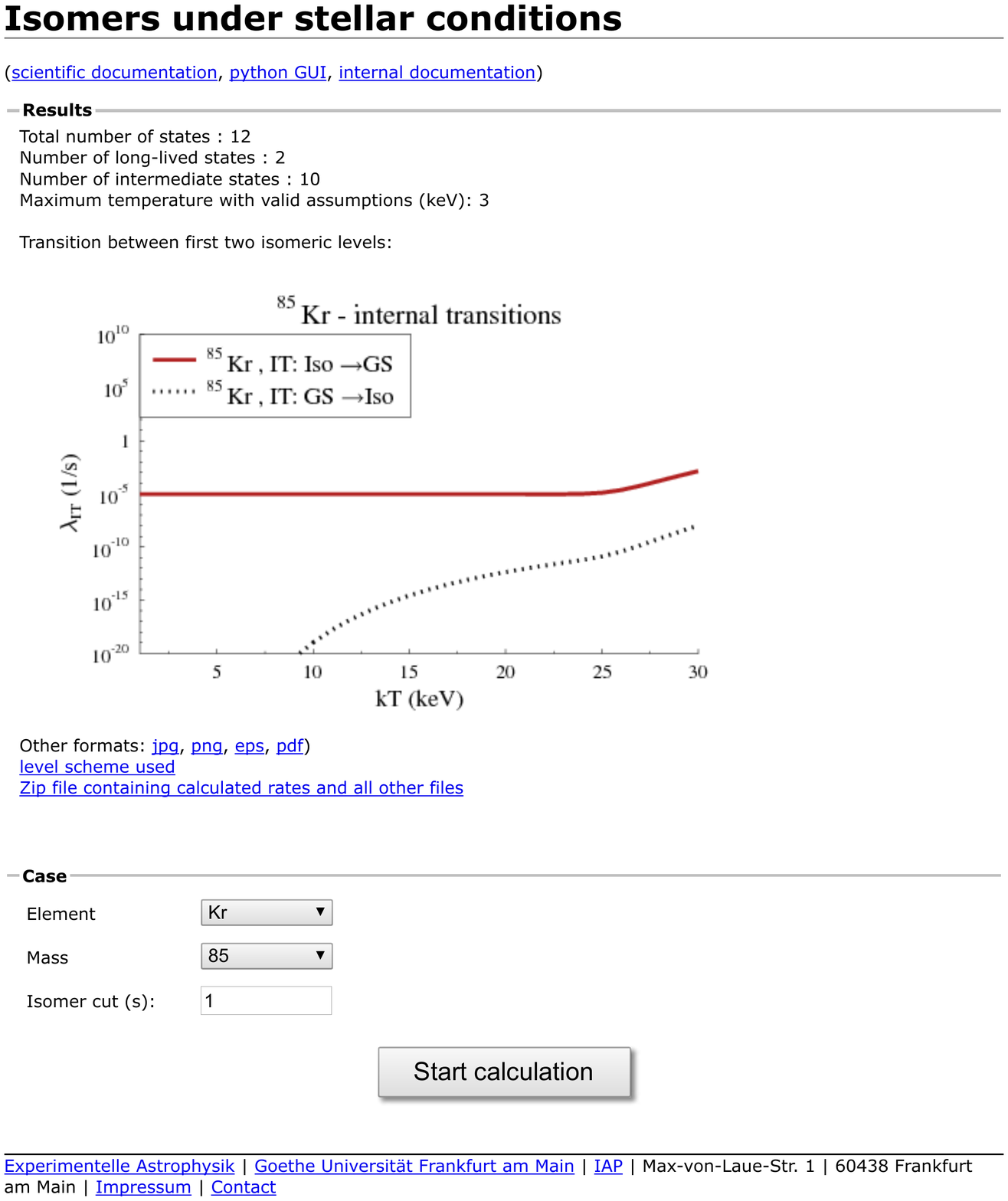}
       \caption{The online interface to calculate and show the coupling rates between long-lived states of selected isomers. The screen shot shows the example of $^{85}$Kr.
               }\label{isomers_online}
      \end{center}
      \end{figure}

\begin{figure}[h]
      \begin{center}
       \includegraphics[width=.999\textwidth]{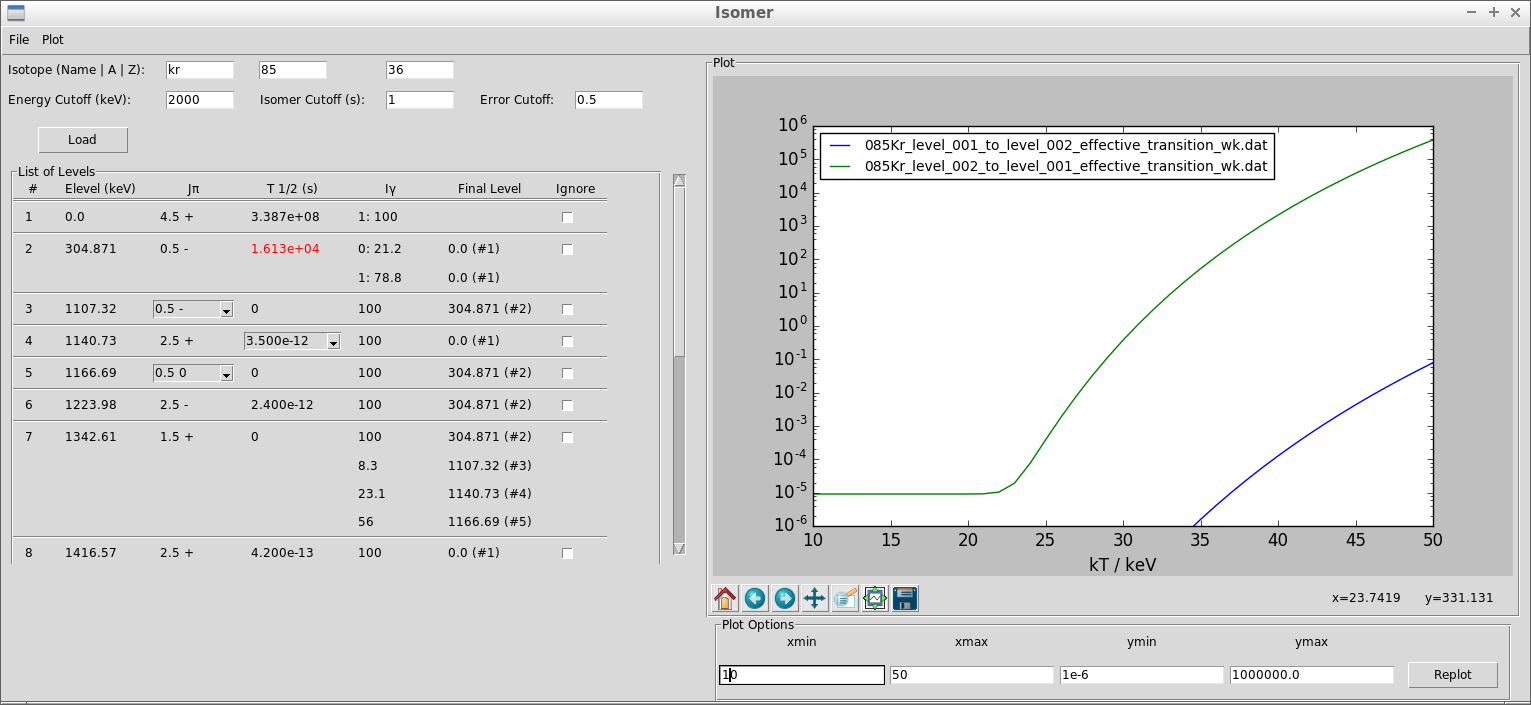}
       \caption{The python interface available to conveniently create the input files necessary to run the code
       calculating the stellar rates. The screen shot shows the example of $^{85}$Kr.
               }\label{interface}
      \end{center}
      \end{figure}

Under the URL http://exp-astro.de/isomers/ we provide a web-interface to the code (Fig.~\ref{isomers_online}) and access to a python interface, which significantly speeds up the process of developing the necessary input files based on the latest nuclear data available (Fig.~\ref{interface}). 

In case of missing nuclear data - usually transitions probabilities between short-lived states - we implemented the Weisskopf-approximation \cite{WeW30}:

\begin{eqnarray}
 \Gamma_{\mathrm{W}(\mathrm{E}l)} & = & \hbar \frac{2 (l+1)}{l \left[(2l+1)!!\right]^2} 
 \left(\frac{3}{l+3}\right)^2 \left(\frac{E_\gamma}{\hbar c}\right)^{2l+1} \alpha c R^{2 l} \\
 \Gamma_{\mathrm{W}(\mathrm{M}l)} & = & \hbar \frac{20 (l+1)}{l \left[(2l+1)!!\right]^2} 
 \left(\frac{3}{l+3}\right)^2 \left(\frac{E_\gamma}{\hbar c}\right)^{2l+1} \alpha c R^{2 l-2} 
 \left(\frac{\hbar}{M_p c}\right)^{2} \\
 t_{1/2} & = & \frac{\hbar \ln{2}}{\Gamma_{\mathrm{W}}}
\end{eqnarray}

Where E$l$ and M$l$ stands for the multipolarity of the transition, $c$ for the speed of light, $\hbar$ for Planck's constant, $M_p$ for the mass of the proton and $\alpha$ for the fine-structure constant. The nuclear radius $R=1.2\cdot A^{1/3}~$fm scales with the number of nucleons $A$ in the nucleus.

\clearpage

\section{Effective live times of isotopes}

In many cases, only two long-lived states are important. If the half-life of these states are very different, it is sometimes useful to introduce a half-life, which is valid for the sum of the 2 species. With $N_s$ the number of nuclei in the short(er)-lived isomeric state and $N_l$ the number of nuclei in the long-lived or even stable state, we are looking for a decay rate with:

\begin{equation}
  N_s(t)+N_l(t) \approx \mbox{e}^{-\lambda_\mathrm{isotope} t}
\end{equation}

There are basically two different temperature regimes. 
\begin{itemize}
	\item{At low temperatures, at least one of the decay rates $\Lambda_{\beta_i}(T)$ is much faster than the coupling rates between the states $\Lambda_{i \rightarrow j}(T)$. Then one can assume that the decay is determined 
	by how quickly $N_l$ is destroyed via 2 possible decay branches:}

	\begin{equation}
		\lambda_\mathrm{low}(T) \approx \Lambda_{\beta_l}(T) + \frac{\Lambda_{l \rightarrow s}(T) \Lambda_{\beta_s}(T)}{\Lambda_{\beta_s}(T)+\Lambda_{s \rightarrow l}}
	\end{equation}

	\item{At high temperatures, the coupling rates between the states $\Lambda_{i \rightarrow j}(T)$ dominate and determine the equilibrium between the long-lived states:
	
	\begin{equation}
		\Lambda_{l \rightarrow s} N_l \approx \Lambda_{s \rightarrow l} N_s
	\end{equation}
	
	with 
	
	\begin{equation}
		\alpha:=\frac{N_s}{N_l} = \frac{\Lambda_{l \rightarrow s}}{\Lambda_{s \rightarrow l}} 
	\end{equation}
	
	follows:
	
	\begin{equation}
		\lambda_\mathrm{high}(T) \approx \frac{\alpha \Lambda_{\beta_s}(T) + \Lambda_{\beta_l}(T)}{1+\alpha}
	\end{equation}

	}
\end{itemize}

Both approximations fail in the intermediate temperature regime. The actual rate will be lower than both of them, the differential equations need to be solved and the equilibrium condition found. However, for the sake of simplicity, one can just assume:

	\begin{equation}
		\lambda_\mathrm{isotope}(T) \approx \mbox{Minimum}\left(\lambda_\mathrm{high}(T),\lambda_\mathrm{low}(T)\right)
	\end{equation}

and finally the effective half-life of the isotope:

	\begin{equation}
		t_{1/2}(T) = \frac{\ln(2)}{\lambda_\mathrm{isotope}(T)}
	\end{equation}

The exact solution would start with the system of differential equations for the ground (g) and isomeric state (i),

\begin{eqnarray}
\frac{\mathrm{d}N_\mathrm{g}}{\mathrm{d}t}=(-\lambda_\mathrm{gi}-\lambda_{\mathrm{g}_\beta}) \cdot N_\mathrm{g} + \lambda_\mathrm{ig}\cdot N_\mathrm{i} \\
\frac{\mathrm{d}N_\mathrm{i}}{\mathrm{d}t}= \lambda_\mathrm{gi}\cdot N_\mathrm{g} + (-\lambda_\mathrm{ig}-\lambda_{\mathrm{i}_\beta}) \cdot N_\mathrm{i} 
\end{eqnarray}

which can be written in matrix form

\begin{eqnarray}
A \cdot \bm{N}=\bm{\mathrm{d}N}
\end{eqnarray}

with

\begin{eqnarray}
A=\left( \begin{array}{cc}
-\lambda_\mathrm{gi}-\lambda_{g_\beta} & \lambda_\mathrm{ig}\\
\lambda_\mathrm{gi} & -\lambda_\mathrm{ig}-\lambda_{i_\beta} 
\end{array} \right),
\end{eqnarray}
\begin{eqnarray}
\bm{N}=\left(\begin{array}{c}
N_\mathrm{g}\\
N_\mathrm{i}
\end{array} \right),
\bm{\frac{\mathrm{d}N}{\mathrm{d}t}}=\left(\begin{array}{c}
\frac{\mathrm{d}N_\mathrm{g}}{\mathrm{d}t}\\
\frac{\mathrm{d}N_\mathrm{i}}{\mathrm{d}t}
\end{array} \right)
\end{eqnarray}

Once the eigenvalues $\lambda_1,\lambda_2$ and the respective eigenvectors $v_1,v_2$ of $A$ are calculated, the general solution for a system of two linear differential equations is used:

\begin{eqnarray}
\bm{x(t)}= c_1\cdot e^{\lambda_1 t}\bm{v_1}+c_2\cdot e^{\lambda_2 t}\bm{v_2}
\end{eqnarray}

The constants $c_1, c_2$ are derived from the initial abundance vector

\begin{eqnarray}
\bm{N_0}=\left(\begin{array}{c}
N_{\mathrm{g}_0}\\
N_{\mathrm{i}_0}
\end{array} \right)
\end{eqnarray}

by solving the system of linear equations

\begin{eqnarray}
\left(\begin{array}{cc}
\bm{v_1} & \bm{v_2} 
\end{array} \right) 
\left(\begin{array}{c}
c_1\\
c_2
\end{array} \right)=\bm{N_0}
\end{eqnarray}

Since $\lambda_\mathrm{gi},\lambda_\mathrm{ig},\lambda_{\mathrm{g}_\beta},\lambda_{\mathrm{i}_\beta}$ are temperature dependent values, the system above needs to be solved independently for each temperature in order to determine the temperature dependent effective lifetime of the isotope. The final result is the abundance vector

\begin{eqnarray}
\bm{x}(T,t)= c_1(T)\cdot e^{\lambda_1(T) t}\bm{v_1}(T)+c_2(T)\cdot e^{\lambda_2(T) t}\bm{v_2}(T)
\end{eqnarray}

The code calculates and provides an effective half-life based on the assumptions made here. It is important to emphasize that those assumptions are only valid if:

\begin{itemize}
  \item Only two long-lived states exist.
  \item The production of either of the two states from other isotopes can be neglected.
\end{itemize}

\section{Some examples of astrophysical interest}

\subsection{The observation novae via $\gamma$-rays - $^{34}$Cl}

\begin{figure}[h]
      \begin{center}
       \includegraphics[width=.7\textwidth]{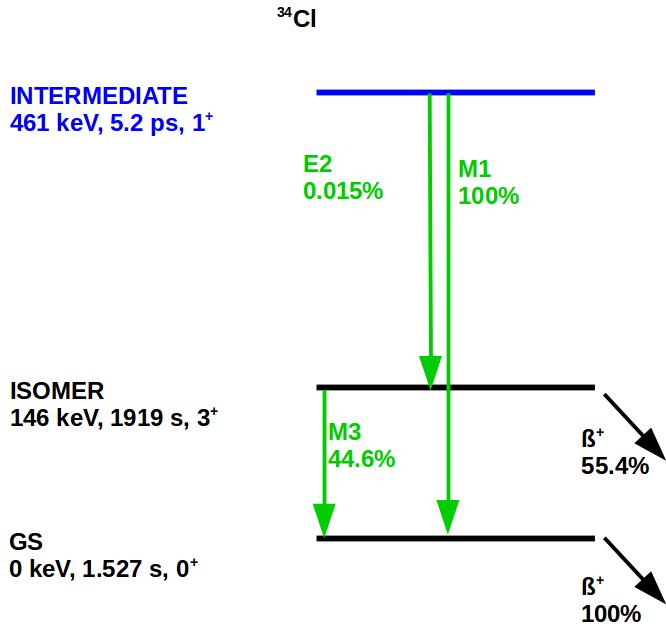}
       \caption{Simplified level scheme of $^{34}$Cl. Higher-lying levels do not contribute below $kT\approx100~$keV. The most important and largely unknown nuclear physics input is the transition from the $1^+$ intermediate state at 461~keV to the isomer at 146~keV. Only an upper limit is known experimentally.
               }\label{cl34_scheme}
      \end{center}
      \end{figure}

A first example is given in Fig.~\ref{cl34_scheme} for the case of $^{34}$Cl, which has a rather short-lived ground state ($t_{1/2}=1.5$~s) and an isomer at 146~keV  ($t_{1/2}=32$~min). The intermediate $1^+$ level at 461~keV acts as the coupling state. This isotope and the corresponding rates are important during novae, since the delayed $\gamma$-rays can potentially be detected \cite{CPN00}.

\begin{figure}[h]
      \begin{center}
       \includegraphics[width=.8\textwidth]{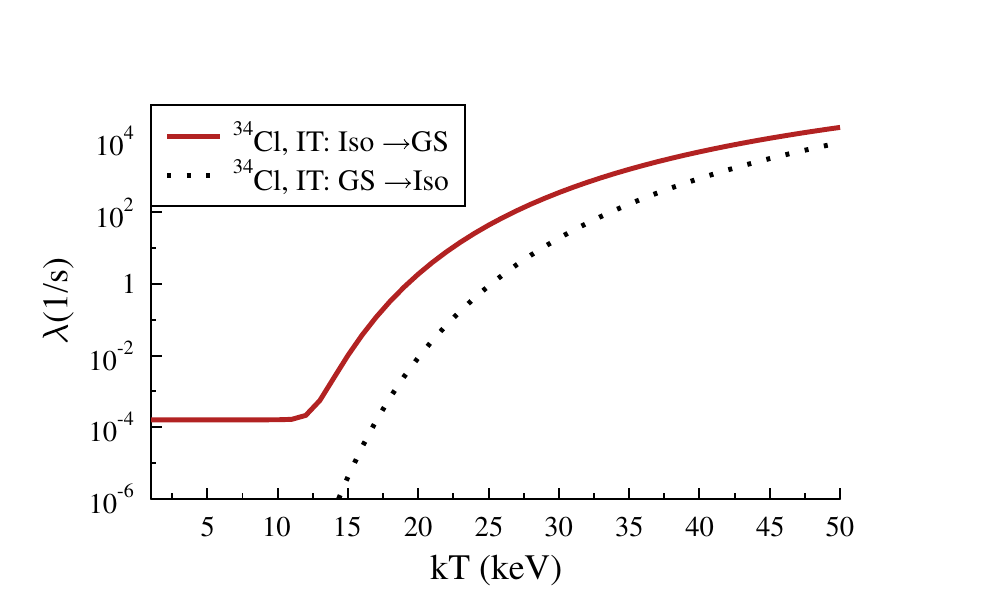}
       \caption{Effective rates coupling the two long-lived states of $^{34}$Cl under stellar conditions
                as a function of the temperature. Two long-lived and one short-lived state were considered.
               }\label{show_cl34_rates12}
      \end{center}
      \end{figure}

It is important to emphasize that the coupling can only occur, if a value for the transition from the intermediate state to the isomer at 146~keV is assumed. Only an upper limit of 0.5\% is experimentally known for this transition. If the transition is left empty in the input file, the code automatically estimates the transition strength based on the Weisskopf approach to 0.022\%. Fig.~\ref{show_cl34_rates12} shows the resulting coupling between the two long-lived states. The consideration of higher-lying states does not change the result at all below $kT\approx100~$keV.

\begin{figure}[h]
      \begin{center}
       \includegraphics[width=.8\textwidth]{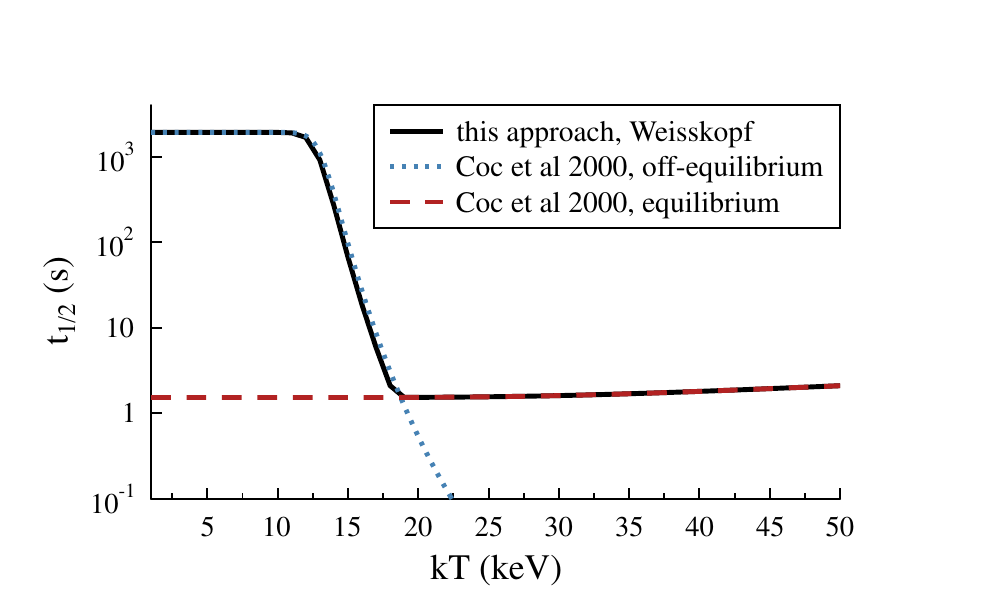}
       \caption{The approximate estimate of the effective half lifetime of $^{34}$Cl under stellar conditions
                as a function of the temperature. Here, no value for the transition to the isomer was set in the input files. The transition probability was automatically calculated using the Weisskopf approximation. A comparison with the estimate by Coc $et~al.$ \cite{CPN00} shows very good agreement.
               }\label{show_cl34_ratesbw}
      \end{center}
      \end{figure}

\begin{figure}[h]
      \begin{center}
       \includegraphics[width=.8\textwidth]{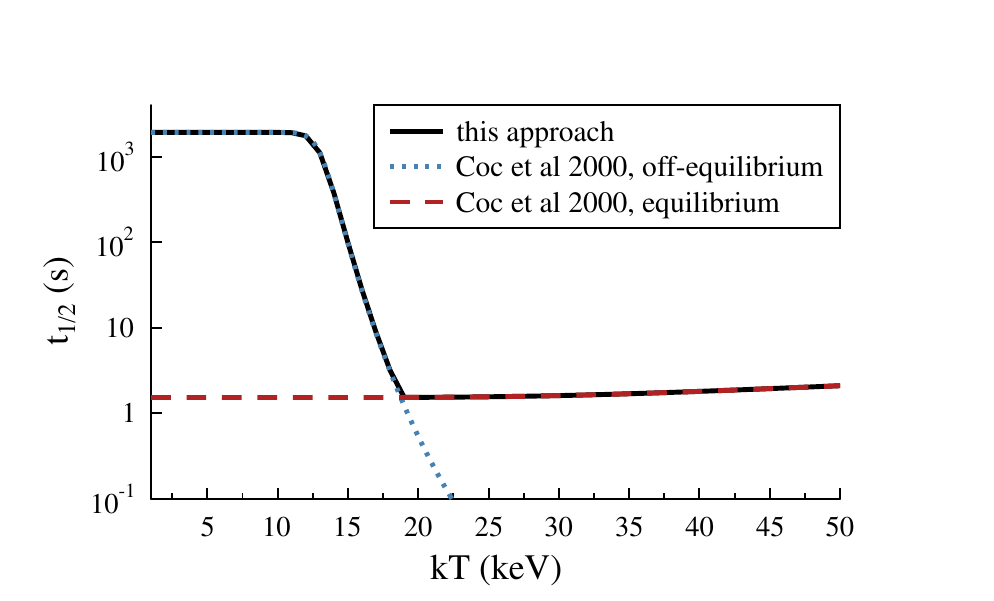}
       \caption{The approximate estimate of the effective half-life of $^{34}$Cl under stellar conditions
                as a function of the temperature and a comparison with previously published estimates \cite{CPN00}. 
               }\label{show_cl34_ratesbo}
      \end{center}
      \end{figure}

Our general approach to the problem reproduces the results of the very sophisticated estimates of Coc $et~al.$ nicely \cite{CPN00}, Fig.~\ref{show_cl34_ratesbw}. Perfect agreement is reached if the transition probability of the intermediate state to the isomer is set to 0.015\% (Fig.~\ref{show_cl34_ratesbo}). Both values are well below the rather large experimental upper limit.

\clearpage

\subsection{Observation of nucleosynthesis in massive stars - $^{26}$Al}

The second interesting case is $^{26}$Al, which can be observed in the center of the milky way via the emitted $\gamma$-rays \cite{DDB95}. In particular the simultaneous observation of the decay of $^{60}$Fe is considered as a clear signature of ongoing stellar nucleosynthesis and allows constraints on our models of massive stars \cite{TWH95,LiC06}.

\begin{figure}[h]
      \begin{center}
       \includegraphics[width=.7\textwidth]{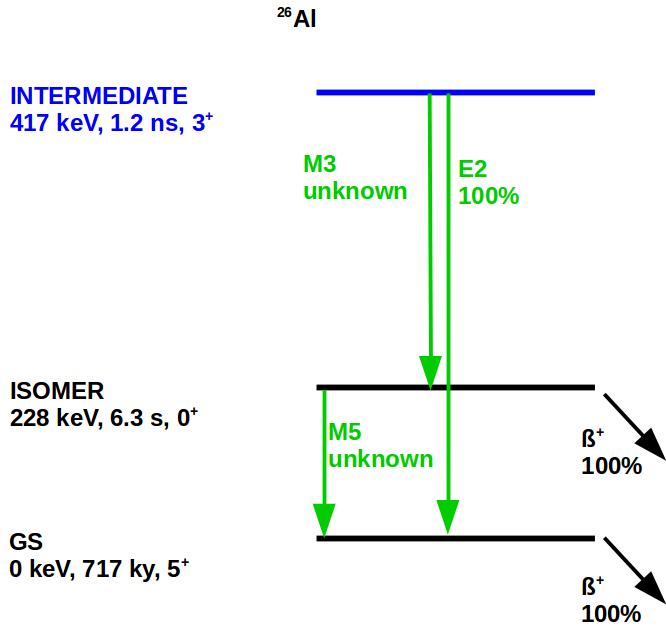}
       \caption{Simplified level scheme of $^{26}$Al. Eight additional, higher-lying levels are considered during the rate calculations, but not shown here. 
               }\label{al26_scheme}
      \end{center}
      \end{figure}

$^{26}$Al has a very long-lived ground state with about 720'000 years of half life and an isomeric state at 228~keV with a half life of 6.3~s (Fig.~\ref{al26_scheme}). Both states decay with almost 100\% probability via electron capture to $^{26}$Mg. However, the unknown M5 transition of the isomer to the ground state needs to be considered. Based on the Weisskopf approximation it is $5\cdot10^{-13}$\%.
In this particular case, the 1.2~ns isomer at 417~keV is very important. It does not need to be treated as an isomer, but the experimentally undetermined M3 transition to the long-lived isomeric state at 228~keV needs to be taken into account. Based on the Weisskopf approximation, it is $1.4\cdot10^{-9}$\%.

\begin{figure}[h]
      \begin{center}
       \includegraphics[width=.8\textwidth]{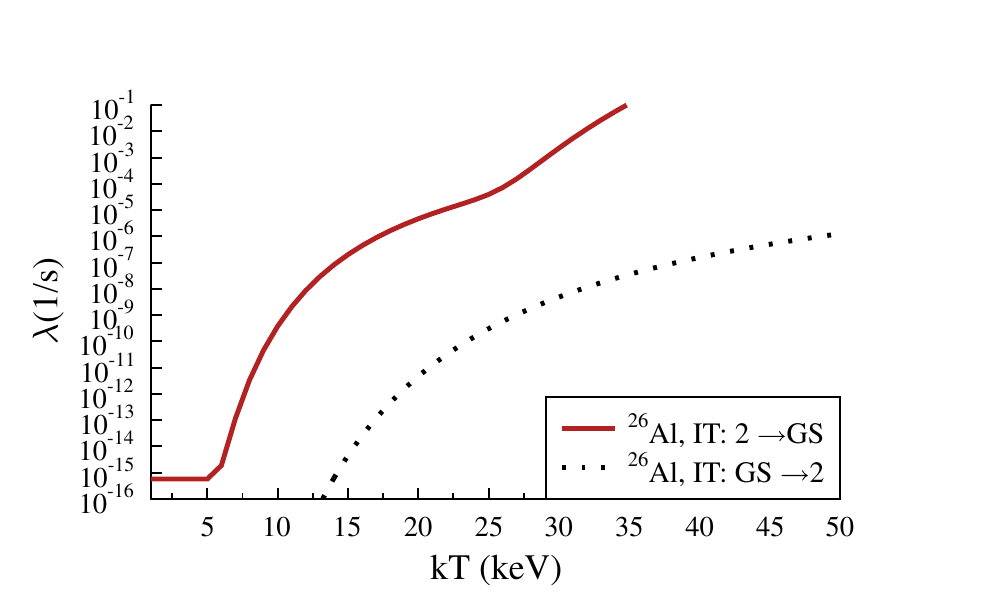}
       \caption{Effective rates coupling the two long-lived states of $^{26}$Al under stellar conditions as a function of the temperature.
               }\label{show_al26_rates12}
      \end{center}
      \end{figure}

The decision, which states should be treated as isomers and which as intermediate states, can be easily made in the input files. A parameter will be passed on to the rate-calculation program, which acts as the separation between long- and short-lived states. States with half lives longer than this parameter will be treated as long-lived states. Fig.~\ref{show_al26_rates12} illustrates the results obtained with a cut of 1~s, which means that only the long-lived isomer and the ground state will be treated explicitly. In addition to the three levels discussed so far, 8 higher-lying levels were considered too.

\begin{figure}[h]
      \begin{center}
       \includegraphics[width=.8\textwidth]{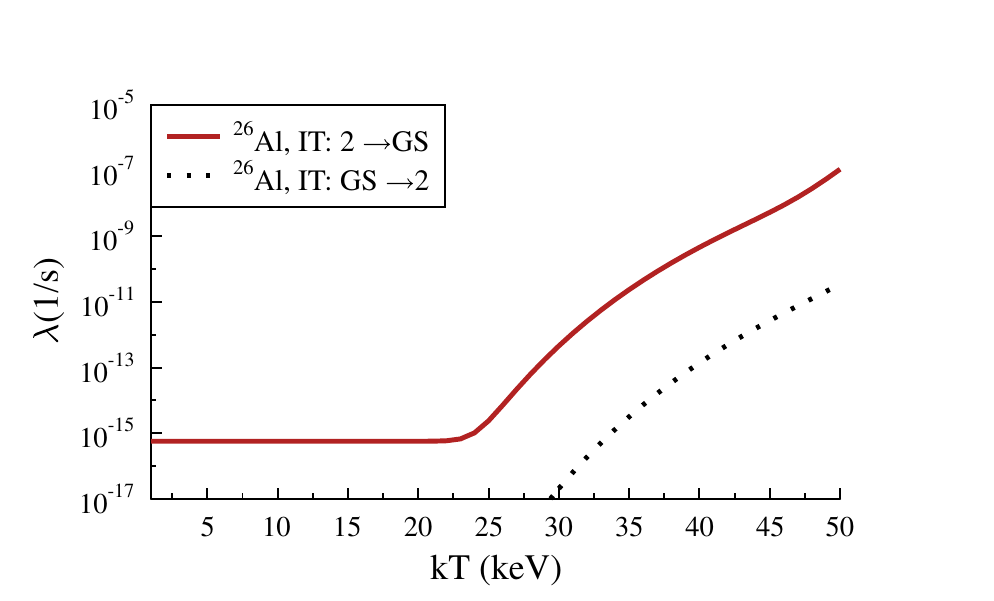}
       \caption{Coupling rates of the ground state and the long-lived isomer (second level in Fig.~\ref{al26_scheme}) of $^{26}$Al under stellar conditions as a function of the temperature. The short-lived isomer (third  level in Fig.~\ref{al26_scheme}) was treated explicitly. Above $kT\approx5$~keV strong deviations between the coupling rates of the pure 2-state treatment are visible (see Fig.~\ref{show_al26_rates12}). 
               }\label{show_al26_1ns_1_2}
      \end{center}
      \end{figure}

\begin{figure}[h]
      \begin{center}
       \includegraphics[width=.8\textwidth]{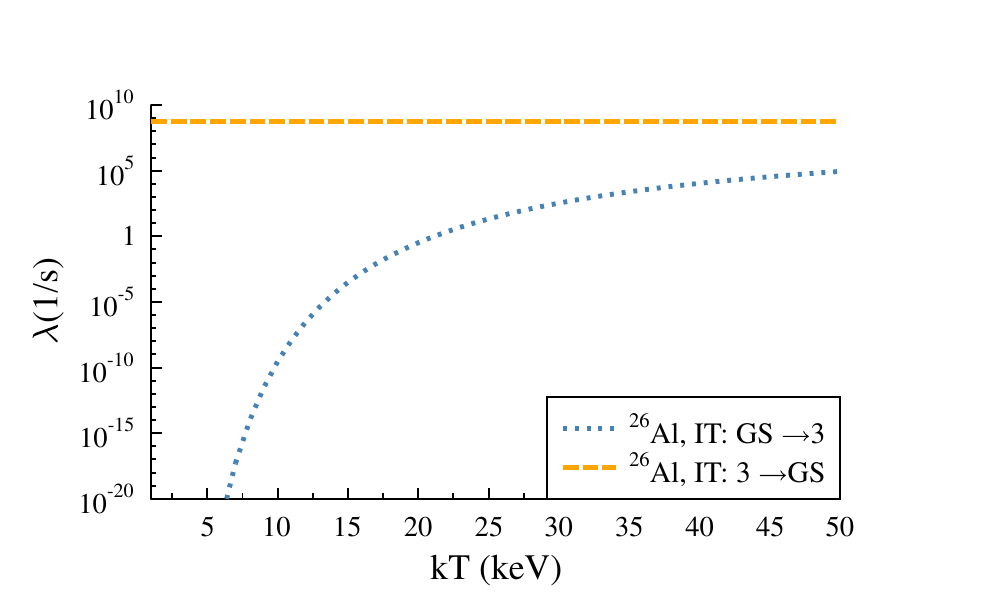}
       \caption{Coupling rates of the ground state and the short-lived isomer (third  level in Fig.~\ref{al26_scheme}) of $^{26}$Al under stellar conditions as a function of the temperature. The short-lived isomer was treated explicitly. 
               }\label{show_al26_1ns_1_3}
      \end{center}
      \end{figure}

\begin{figure}[h]
      \begin{center}
       \includegraphics[width=.8\textwidth]{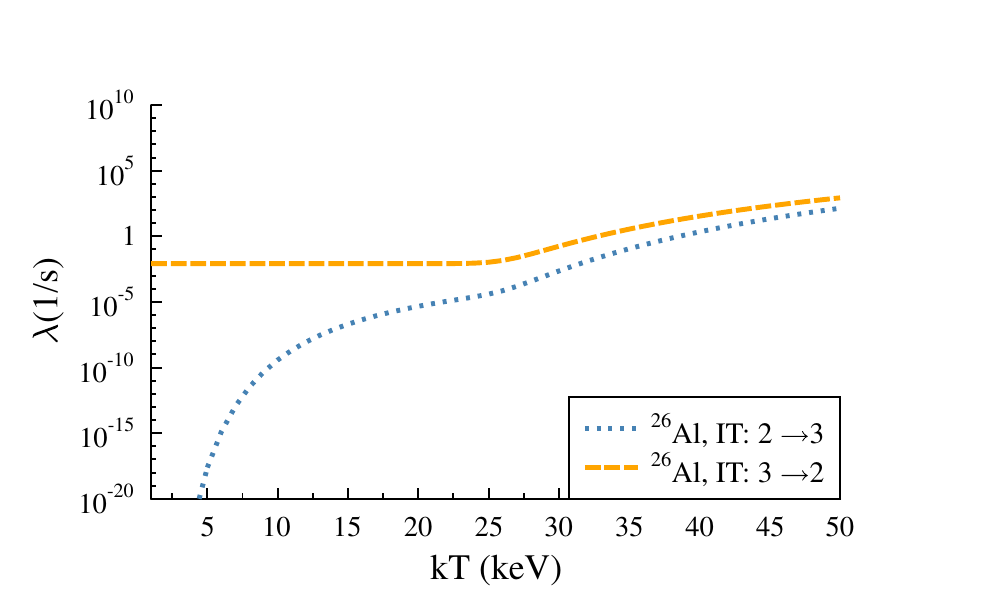}
       \caption{Coupling rates of the long-lived isomer (second  level in Fig.~\ref{al26_scheme}) and the short-lived isomer (third  level in Fig.~\ref{al26_scheme}) of $^{26}$Al under stellar conditions as a function of the temperature. The short-lived isomer was treated explicitly. 
               }\label{show_al26_1ns_2_3}
      \end{center}
      \end{figure}

It is very illustrative to set the isomer cut to 1~ns. In this case, 3 states are treated explicitly and 3 combinations of transitions exist. Since the transition between the long-lived isomer and the ground state occurs at high temperatures mostly via the short-lived isomer, the rates coupling ground state and isomeric state are very different depending on the explicit or implicit treatment of the short-lived isomer, compare Figs.~\ref{show_al26_rates12} and \ref{show_al26_1ns_1_2}. The rates between the two states differ significantly above $kT\approx5$~keV. The explanation can be found in Figs.~\ref{show_al26_1ns_1_3} and \ref{show_al26_1ns_2_3}. The short-lived isomer starts to become thermally excited from ground state and long-lived isomer at $kT\approx5$~keV.

However, if the stellar conditions (temperatures, production conditions) are changing on a scale slower the scale of 1~ns, the nucleosynthesis results will be the same - no matter which implementation was chosen.

\begin{figure}[h]
      \begin{center}
       \includegraphics[width=.8\textwidth]{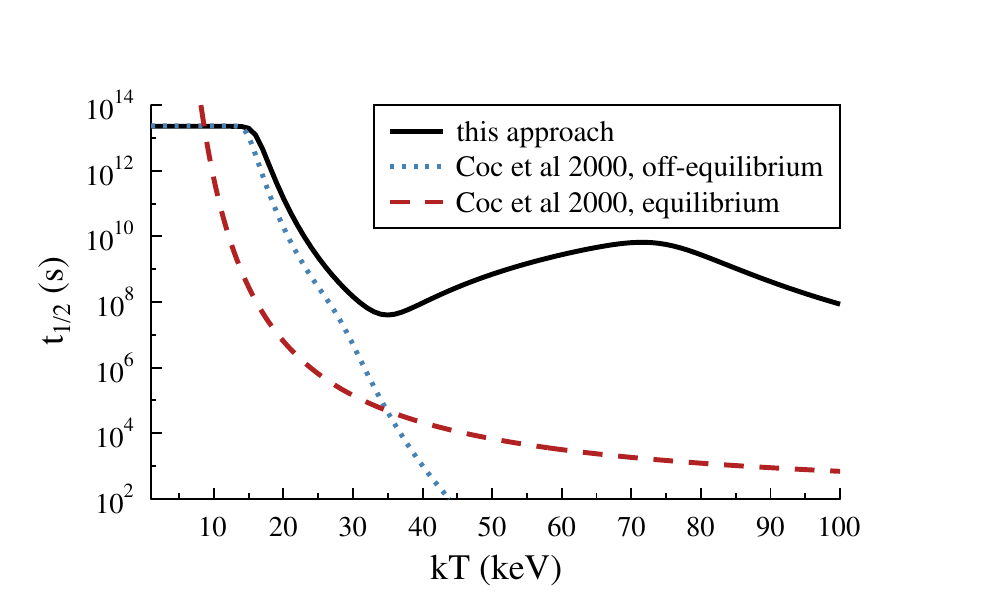}
       \caption{The approximate estimate of the effective half-life of $^{26}$Al under stellar conditions as a function of the temperature. Here, no value for the transition to the isomer was set in the input files. The missing transition probabilities were automatically calculated using the Weisskopf approximation. Above 30~keV our estimate differs significantly from a previous estimate by Coc $et~al.$ \cite{CPN00}.
               }\label{show_al26_ratesbw}
      \end{center}
      \end{figure}

\begin{figure}[h]
      \begin{center}
       \includegraphics[width=.8\textwidth]{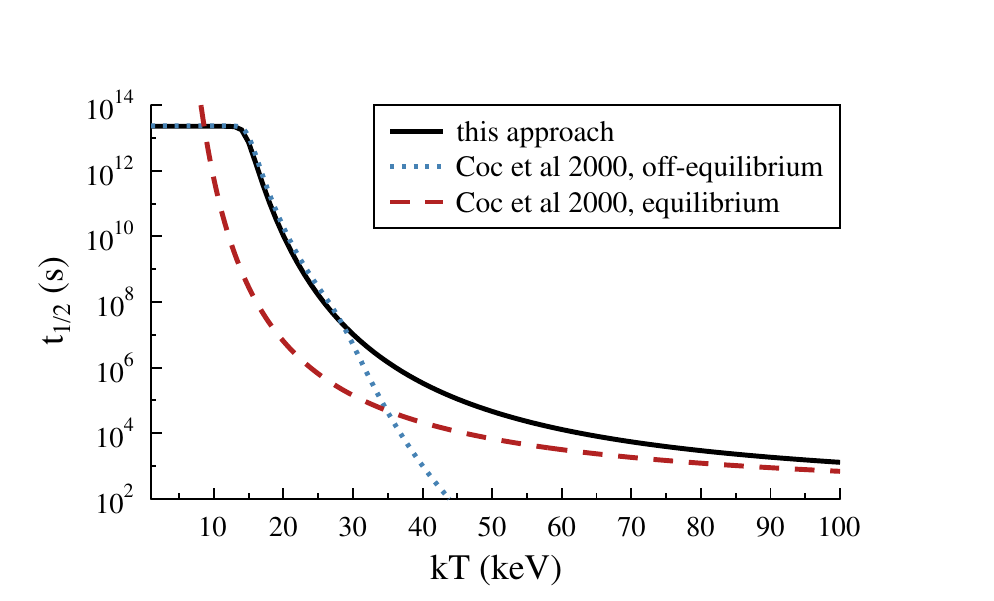}
       \caption{The approximate estimate of the effective half-life of $^{26}$Al under stellar conditions as a function of the temperature. Only the first 3 states (Fig.~\ref{al26_scheme}) are considered. These are basically the assumptions made by Coc $et~al.$ \cite{CPN00}.
               }\label{show_al26_ratesbo}
      \end{center}
      \end{figure}

Fig.~\ref{show_al26_ratesbw} shows a comparison of the derived effective half-life of $^{26}$Al under stellar conditions with a previous estimate \cite{CPN00}. The two estimations differ significantly above $kT\approx30$~keV. We found that the main reason for the deviation are higher-lying states. While Coc $et~al.$ considered only the first 4 states, we considered the first 11 states. The huge difference is remarkable, in particular since the 4th state has already an excitation energy of 1.1~MeV and the 5th at 1.76~MeV. Fig.~\ref{show_al26_ratesbo} shows the outcome of our code, if only the first 3 states are considered and the transition rates of the 1-ns-isomer are slightly adjusted. Obviously, the Coc $et~al.$ estimates can be reproduced, if similar assumptions are made.

\clearpage

\subsection{Dating the universe - $^{85}$Kr}

\begin{figure}[h]
      \begin{center}
       \includegraphics[width=.7\textwidth]{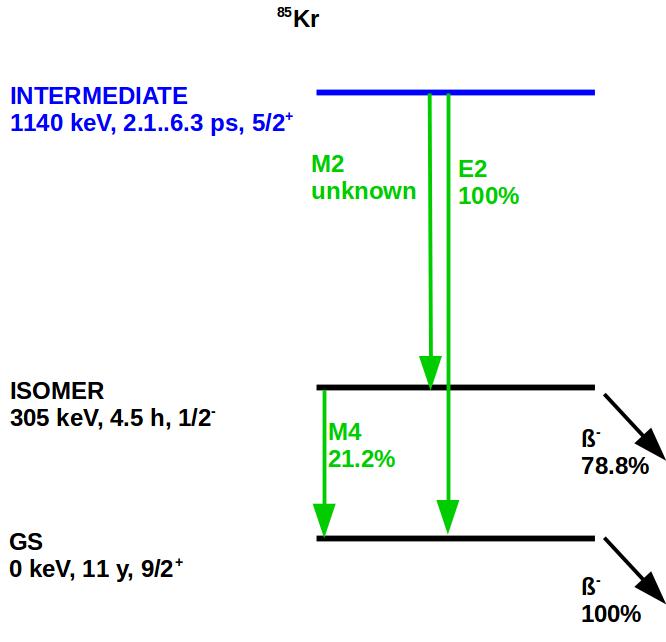}
       \caption{Simplified level scheme of $^{85}$Kr. Only one representative intermediate level out of 10 considered levels is shown here. This level is also a representative example of the poor knowledge of the nuclear properties of the excited levels.
               }\label{kr85_scheme}
      \end{center}
      \end{figure}

The s-process branching at $^{85}$Kr determines the astrophysical observable ratios $^{84}$Kr/$^{86}$Kr \cite{PGA06} as well as $^{87}$Rb/$^{87}$Sr, which can be used as a cosmo-chronometer \cite{BeW84c}. The correct treatment of the isomeric state at 305~keV is extremely important for the understanding of the nucleosynthesis outcome. The ground state of  $^{85}$Kr has a half life of about 10~yr and the isomeric state about 4.5~h. The isomer can decay to the ground state, Fig.~\ref{kr85_scheme}.

\begin{figure}[h]
      \begin{center}
       \includegraphics[width=.8\textwidth]{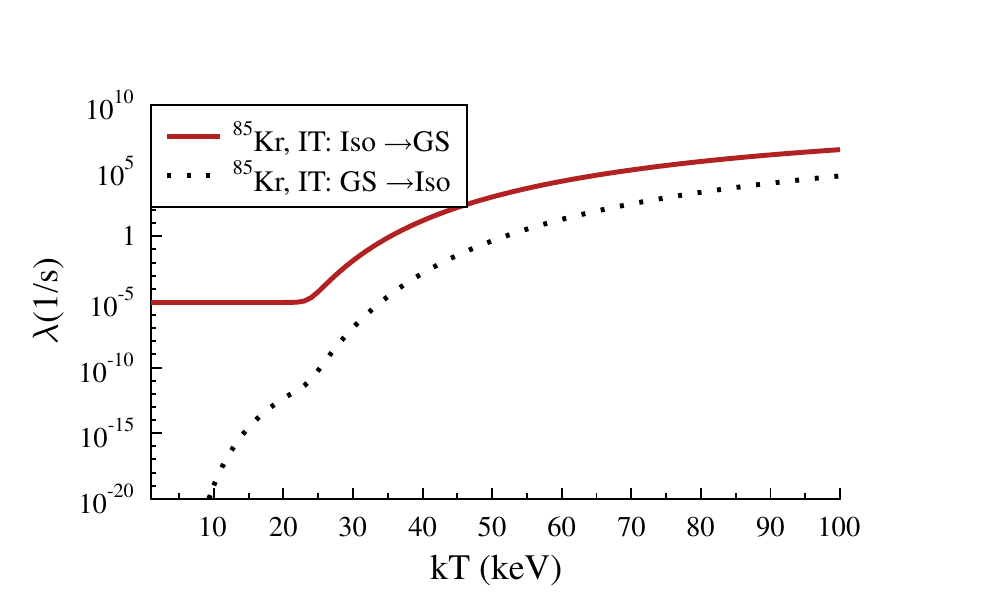}
       \caption{Effective rates coupling the two long-lived states of $^{85}$Kr under stellar conditions as a function of the temperature. 
               }\label{show_kr85_rate12}
      \end{center}
      \end{figure}

The transition of the isomer to the ground state will be thermally enhanced if higher-lying states can be thermally populated. Fig.~\ref{show_kr85_rate12} shows the result if the first 12 levels of $^{85}$Kr are considered. The knowledge of the nuclear properties, however, is rather poor. Better data on the live times, angular momentum and parity of the higher-lying states are urgently needed. 

\begin{figure}[h]
      \begin{center}
       \includegraphics[width=.8\textwidth]{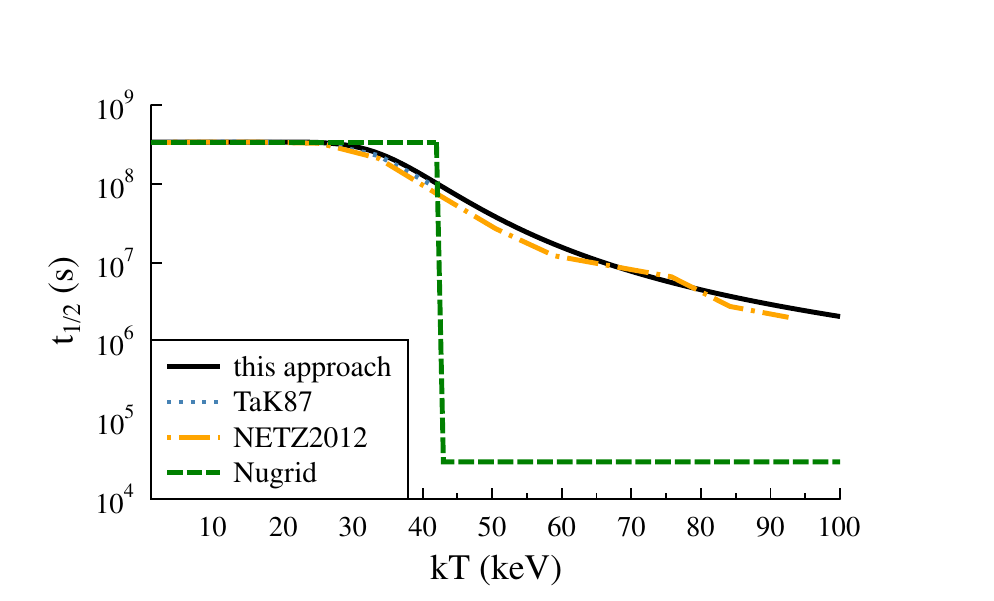}
       \caption{The approximate estimate of the effective half-life of $^{85}$Kr under stellar conditions as a function of the temperature with a comparison of realizations in different nucleosynthesis codes \cite{TaY87,Jaa91,WBC15,KTG16}. 
               }\label{show_kr85_rateb}
      \end{center}
      \end{figure}

Fig.~\ref{show_kr85_rateb} shows our estimate of the effective half life under stellar conditions, provided no production and other destruction channels are possible. We compared our results with different implementations into stellar nucleosynthesis codes. For temperatures below 30~keV all implementations agree since no coupling occurs. 

\clearpage

\section{Summary}

We developed a general algorithm to calculate the coupling rates of long-lived states under stellar conditions. The only input parameters are the nuclear properties of the excited states, which can be measured under terrestrial conditions. The algorithm is implemented in a program, which is available online at the URL ''http://exp-astro.de/isomers/''. Currently about a dozen isotopes of astrophysical interest are available. More isotopes will continuously be implemented. Please contact the authors if a particular isotope is of interest and missing.

\section*{Acknowledgments}

This research has received funding from the European Research Council under the European Unions's Seventh Framework Programme (FP/2007-2013) / ERC Grant Agreement n. 615126.

\appendix

\bibliographystyle{ws-ijmpa} 
\bibliography{refbib}

\end{document}